\documentclass[oneside,reqno,12pt]{amsart}

\usepackage{epsfig}
\usepackage{rangecite}
\usepackage{euscript}

\newcommand{\mathscript}{\EuScript}

\numberwithin{equation}{section}

\newcommand{\du}[3]{{\/{#1}_{#2}}^{#3}}
\newcommand{\ud}[3]{{\/{#1}^{#2}}_{#3}}

\newcommand{\ald}{{\dot{\alpha}}}

\newcommand{\bed}{{\dot{\beta}}}

\newcommand{\Ad}{A^*}
\newcommand{\Bd}{B^*}
\newcommand{\alb}{{\underline{\alpha}}}
\newcommand{\beb}{{\underline{\beta}}}
\newcommand{\gab}{{\underline{\gamma}}}
\newcommand{\vev}[1]{\left\langle #1 \right\rangle}
\newcommand{\Pdag}{\Phi^{\dag}}
\newcommand{\sbar}{\bar{\sigma}}
\newcommand{\pb}{\bar{\psi}}
\newcommand{\D}{\mathscript{D}}
\newcommand{\Lag}{\mathscript{L}}
\newcommand{\E}{\mathscript E}
\newcommand{\proj}{(\bar{\D}^2 - 8 R)}

\newcommand{\tb}{\bar{\theta}}

\newcommand{\bR}{\bar{\mathscript{R}}}
\newcommand{\bn}{\bar{\nabla}}

\newcommand{\R}{\mathscript{R}}

\makeatletter
    \def\serieslogo@{\vtop to 0pt{\noindent\scriptsize\ppn\parindent\z@}}
    \let\@setcopyright\@empty
\makeatother

\begin{document}

\def\ppn{HEP-TH/9511223, UPR-685T}
\title[SUSY Breaking in Four-Dimensional Supergravity]{Four-Dimensional 
Higher-Derivative Supergravity \\ and Spontaneous Supersymmetry 
Breaking}
\author{Ahmed Hindawi, Burt A. Ovrut, and Daniel Waldram}
\thanks{Published in Nuclear Physics B \textbf{476} (1996), 175--199.}
\maketitle
\vspace*{-0.3in}
\begin{center}
\small{\textit{Department of Physics, University of Pennsylvania}} \\
\small{\textit{Philadelphia, PA 19104-6396, USA}}
\end{center}

\begin{abstract}

We construct two classes of higher-derivative supergravity theories 
generalizing Einstein supergravity. We explore their dynamical content 
as well as their vacuum structure. The first class is found to be 
equivalent to Einstein supergravity coupled to a single chiral 
superfield. It has a unique stable vacuum solution except in a special 
case, when it becomes identical to a simple no-scale theory. The second 
class is found to be equivalent to Einstein supergravity coupled to two 
chiral superfields and has a richer vacuum structure. It is demonstrated 
that theories of the second class can possess a stable vacuum with 
vanishing cosmological constant that spontaneously breaks supersymmetry. 
We present an explicit example of this phenomenon and compare the result 
with the Polonyi model.

\vspace*{\baselineskip}


\end{abstract}

\renewcommand{\baselinestretch}{1.2} \large \normalsize

\vspace*{\baselineskip}

\section{Introduction}

In a recent series of papers \cite{PRD-53-5583,PRD-53-5597,NPB-471-409}, 
we studied the general structure of higher-derivative gravitational theories, 
both for bosonic gravity and for two-dimensional $(1,1)$ supergravity. Although 
a wide range of technical phenomena, such as consistent coupling of spin-$2$ 
fields to gravity and the method of super-Legendre transformations, were 
explored, the main thesis of these papers lay elsewhere. We systematically 
showed that both bosonic and supergravitational higher-derivative theories 
exhibit a rich structure of non-trivial vacua. These vacua can be expressed as 
non-vanishing vacuum expectation values of the scalar fields that generically 
arise as new degrees of freedom in such theories. While these vacua always have 
non-vanishing cosmological constant in pure bosonic gravity and, hence, are of 
restricted interest in particle physics, we showed that non-trivial vacua with 
zero cosmological constant do arise in two-dimensional supergravity. 
Furthermore, these vacua spontaneously break supersymmetry \cite{NPB-471-409}. 
It follows that higher-derivative supergravity theories can potentially play an 
important role in particle physics, including introducing what is essentially a 
new method of supersymmetry breaking. To realize this goal, however, it is 
essential to extend our results to four-dimensional, $N=1$ supergravity. This 
will be the content of this paper.

In Section~2, we will briefly present generic results in $D=4$, $N=1$ 
supergravity that we will need later in the paper. Section~3 is devoted to a 
short exposition of the methods and results in purely bosonic higher-derivative 
gravitation that we obtained in \cite{PRD-53-5583,PRD-53-5597}. This sets the 
stage for the extension to four-dimensional, $N=1$ supergravity. We present a 
simple higher-derivative, $D=4$, $N=1$ supergravity theory in Section~4 and 
explore its vacuum structure. We show that a subset of such theories is 
equivalent to no-scale supergravity with a single chiral supermultiplet 
\cite{PLB-133-61}. Finally, in Section~5 we give the generic four-dimensional 
higher-derivative, $N=1$ supergravity theory involving the scalar 
curvature superfield $R$ only. We show that these theories contain 
non-trivial vacuum states, with vacuum expectation values of the order 
of the Planck mass $M_P$, that have vanishing cosmological constant and 
that spontaneously break supersymmetry. Furthermore, all fields around 
these vacua propagate physically; that is, there are no ghosts. Finally, 
we then show that in the limit of large $M_P$, these theories become 
equivalent to a generalized type of Polonyi model \cite{PP-KFKI-93}.

These results open the possibility that in phenomenological supergravity 
models and, perhaps, in superstring theories \cite{PLB-388-512}, supersymmetry 
is spontaneously broken not by a hidden sector or by gaugino 
condensates, for example, but simply by the new degrees of freedom that 
arise in higher-derivative terms in the super-effective Lagrangian.

\section{$D=4$, $N=1$ supergravity}

In this section we give a brief review of four-dimensional $N=1$ 
supergravity from the superfield point of view. We will discuss both 
pure supergravitation and supergravity coupled to chiral matter. This is 
done both to set our notation and to present some of the explicit 
formulas we will need later in the paper. Here and elsewhere we follow 
the superfield formulation of supergravity presented in \cite{WB-SS}.

Consider a supermanifold with coordinates 
$z^M=(x^m,\theta^\alpha,\tb_{\dot{\alpha}})$ where $x^m$ are the 
ordinary spacetime commuting coordinates while $\theta^\alpha$ and 
$\tb_{\dot{\alpha}}$ are fermionic anti-commuting coordinates. 
Hereafter, Einstein indices that transform under coordinate 
transformations are denoted by $(M,N,\ldots)$, whereas Lorentz indices 
that transform under the structure group are denoted by $(A,B,\ldots)$.

The geometry of the superspace is determined by the supervielbein $\du 
EMA$ and the Lie algebra valued connection one-form $\du\phi{MA}B$. The 
torsion is defined as the covariant derivative of the supervielbein
\begin{align}
T^A &= dE^A + E^B \du\phi{B}A \notag \\
    &= \tfrac12 dz^M dz^N \du{T}{NM}A. 
\end{align}
The curvature tensor is a Lie algebra valued 2-form defined in terms of 
the connection by
\begin{equation}
R= d\phi + \phi\wedge\phi.
\end{equation}

The number of components in the torsion and curvature is very large and 
a set of constraints is required to reduce it. The old minimal 
supergravity theory is defined by applying the following constraints
\begin{equation}
\begin{split}
\du{T}{\alb\beb}{\gab}&=0, \qquad \du{T}{\alpha\beta}c = 
\du{T}{\ald\bed}c=0, \\
\du{T}{\alpha\bed}c &= \du{T}{\bed\alpha}c=2i\du\sigma{\alpha\bed}c, \\
\du{T}{\alb b}c &= \du{T}{a \beb}c = 0, \\
\du{T}{ab}{c} &= 0,
\end{split}
\label{con}
\end{equation}
where $\alb$ denotes either $\alpha$ or $\ald$. The torsion and 
curvature satisfy the Bianchi identities
\begin{equation}
\begin{split}
\D\D E^A &= E^B \du{R}BA, \\
\D T^A &= E^B \du{R}BA.
\end{split}
\label{BI}
\end{equation}
One has to solve these identities subject to the constraints \eqref{con} 
to find the reduced set of fields. We merely state the results here. It 
turns out that all superfield components of the torsion and curvature 
can be expressed in terms of three superfields, a chiral superfield $R$, 
an hermitian vector superfield $G_{\alpha\dot{\alpha}}$ and a chiral 
superfield $W_{\alpha\beta\gamma}$ totally symmetric in its indices. The 
component fields of the reduced supergravity multiplet are found to be 
the graviton $\du ema(x)$, the gravitino $\du\psi m\alpha(x)$ and two 
additional fields, $M(x)$, a complex scalar, and a real vector field 
$b_m(x)$.

Beside the supergravity multiplet, we need to introduce matter fields. 
We will restrict our discussion to chiral superfields which are 
superfields satisfying the covariant condition
\begin{equation}
\bar{\D}_\ald \Phi = 0.
\end{equation}
Chiral superfields contain three component fields. The expansion of 
chiral superfields in terms of $\theta$ and $\tb$ is complicated and 
coordinate-dependent, since $\theta$ and $\tb$ carry Einstein indices. 
It is better to define the component fields of a chiral superfield 
$\Phi$ by
\begin{equation}
\begin{split}
A &= \Phi|_{\theta=\tb=0}, \\
\chi_\alpha &= \tfrac{1}{\sqrt{2}} \D_\alpha\Phi|_{\theta=\tb=0}, \\
F &= -\tfrac14 \D^\alpha \D_\alpha \Phi|_{\theta=\tb=0}.
\end{split}
\end{equation}
These components carry Lorentz indices. New fermionic coordinates 
$\Theta^\alpha$ are defined such that the expansion coefficients of 
chiral superfields are precisely the covariant components $A$, 
$\chi_\alpha$, and $F$. That is
\begin{equation}
\Phi = A(x) + \sqrt{2} \Theta^\alpha \chi_\alpha(x) + \Theta^\alpha 
\Theta_\alpha F(x).
\end{equation}
These are referred to as chiral superspace coordinates.

We are now in a position to write down the pure supergravity theory as 
well as theories of supergravity coupled to chiral matter. The Einstein 
supergravity Lagrangian is given by
\begin{equation}
\Lag = - \frac{3}{\kappa^2} \int d^4\theta E,
\end{equation}
where $E$ is the super-determinant of the supervielbein and 
$\kappa^2=8\pi G_N$ is the gravitational coupling constant which we will 
set equal to unity unless stated otherwise. This Lagrangian can be 
written as an integral over chiral superspace as
\begin{equation}
\Lag = - 6 \int d^2\Theta \E R + \text{h.c.},
\label{ER}
\end{equation}
where $\E$ and $R$ have the following $\Theta$ expansions
\begin{equation}
\begin{split}
\E = & \tfrac12 e \Big\{ 1 + i \Theta \sigma^a \bar{\psi}_a - 
\Theta\Theta [ M^* + \bar{\psi}_a \bar{\sigma}^{ab} 
\bar{\psi}_b ] \Big\}, \\
R = & - \tfrac16 \Big\{ M + \Theta \big[ \sigma^a \bar\sigma^b 
\psi_{ab} - i \sigma^a \bar\psi_a M + i \psi_b b^b \big] \\
& + \Theta\Theta  \big[ -\tfrac12 \R + i \bar\psi^a 
\bar\sigma^b \psi_{ab} + \tfrac23 MM^* + \tfrac13 b^a b_a - i \du e a 
m \D_m b^a \\
& +\tfrac12 \bar\psi \bar\psi M - \tfrac12 \psi_a 
\sigma^a \bar\psi_c b^c + \tfrac18 \varepsilon^{abcd} 
\big( \bar\psi_a\bar\sigma_b\psi_{cd} + \psi_a \sigma_b 
\bar\psi_{cd}\big) \big] \Big\}.
\end{split}
\label{E&R}
\end{equation}
Here $e$ is the determinant of the vielbein $\du{e}ma$, $\R$ is the 
spacetime curvature scalar, and 
\begin{equation}
\du\psi{mn}\alpha = \tilde{\D}_m \du\psi{n}\alpha - 
\tilde\D_n\du\psi{m}\alpha,
\end{equation}
where
\begin{equation}
\tilde{\D}_n \du\psi{m}\alpha = \partial_n \du\psi{m}\alpha + 
\du\psi{m}\beta \du\omega{n\beta}\alpha, \qquad 
\du\omega{n\beta}\alpha(x) = \du\phi{n\beta}\alpha(z)|_{\theta=\tb=0}.
\end{equation}
The component field Lagrangian for Einstein supergravity can be obtained 
by substituting \eqref{E&R} into \eqref{ER} and performing the 
$\Theta$-integral. The result is
\begin{equation}
e^{-1} \Lag = -\tfrac12 \R - \tfrac13 M M^* + \tfrac13 b^m b_m + 
\tfrac12 \varepsilon^{k\ell mn} \big( \bar{\psi}_k \sbar_\ell 
\tilde{\D}_m \psi_n - \psi_k \sigma_\ell \tilde{\D}_m \pb_n \big).
\label{LE}
\end{equation}
Clearly $M$ and $b_m$ are auxiliary fields and can be eliminated from 
the Lagrangian using their equations of motion. These are given by
\begin{equation}
\begin{split}
M &= 0, \\
b_m &= 0, \\
\end{split}
\end{equation}
respectively. The remaining bosonic part of the Lagrangian is just 
Einstein gravity which describes the propagation of a massless graviton. 
The fermionic part is a kinetic-energy term for the massless gravitino 
$\psi_{m\alpha}$.

We now turn our attention to the coupling of supergravity to 
chiral matter superfields $\Phi_i$. This coupling is determined by two 
functions, the hermitian K\"ahler potential $K(\Phi_i,\Pdag_i)$ and the 
holomorphic superpotential $W(\Phi_i)$. The Lagrangian of Einstein 
supergravity coupled to the chiral fields $\Phi_i$ is given by
\begin{equation}
\Lag = \int d^4\theta E \left\{ - 3 e^{-\frac13 K(\Phi_i,\Pdag_i)} + 
\frac{W}{2R} + \frac{W^{\dag}}{2R^{\dag}} \right\}.
\label{SG+Matter}
\end{equation}
This is an integral over the full superspace. It can be written as an 
integral over chiral superspace of the form
\begin{equation}
\Lag = -\frac14 \int d^2\Theta \E \proj \left\{ - 3 e^{-\frac13K} \right\} 
- \frac14 \int d^2\Theta \E \proj \left\{ 
\frac{W}{2R} + \frac{W^{\dag}}{2R^{\dag}} \right\} + \text{h.c.}
\label{L3}
\end{equation}
Simplification can be achieved using the following fact \cite{NPB-147-105}
\begin{equation}
\int d^2\Theta \E \proj \left( L - L^{\dag} \right) +\text{h.c.} = \text{total 
divergence},
\label{id}
\end{equation}
which holds for an arbitrary superfield $L$. Hence, a spacetime integral 
of this expression can be dropped from any action because it does not 
contribute to the field equations. Applying this to the superpotential 
term in \eqref{L3} yields
\begin{equation}
\Lag = -\frac14 \int d^2\Theta \E \proj \left\{ 
-3e^{-\frac13 K} \right\} + 2 \int d^2\Theta \E  W + \text{h.c.},
\label{SGMC}
\end{equation}
where we have used the fact that $W$ is chiral.

Using the $\Theta$-expansion of $\E$, $R$, and $\Phi_i$, we can evaluate 
the component field Lagrangian. One can show that $M$, $b_m$ and 
$F_i$ are auxiliary; that is, they are not propagating fields. Their 
equations of motion are purely algebraic and can be used to eliminate 
them from the Lagrangian. After doing so, we get a Lagrangian for the 
propagating fields $\du{e}ma$, $\psi_{m\alpha}$, $A_i$, and 
$\du\chi{i}\alpha$ which has a non-canonical coupling of both the 
scalar curvature $\R$ and the gravitino kinetic-energy term to the 
matter fields. This non-canonical coupling can be transformed away by 
Weyl rescaling the vielbein and making a field-dependent redefinition of 
the spinors in an appropriate way. Performing these transformation, we 
get a canonically normalized but complicated component field Lagrangian. 
We will not reproduce this entire Lagrangian here, but rather will refer 
the reader to \cite{WB-SS}. Of special importance in this paper is the 
bosonic part of the Lagrangian. It is found to be
\begin{equation}
e^{-1} \Lag_{\text{Bosonic}} = - \tfrac12 \R + e^{-1} \Lag_{\text{KE-Bosonic}} 
- V(A_i,A_i^*),
\label{LBosonic}
\end{equation}
where the kinetic energy Lagrangian for the matter fields 
$\Lag_{\text{KE-Bosonic}}$ is given by
\begin{equation}
e^{-1} \Lag_{\text{KE-Bosonic}} = - K_{ij^*} \partial^m A^i \partial_m A^{j*}
\label{LKEBosonic}
\end{equation}
and the scalar potential $V(A_i,A_i^*)$ is given by
\begin{equation}
V(A_i,A_i^*) = e^K \Big\{ K^{ij^*} D_iW (D_jW)^* - 3 |W|^2 \Big\},
\label{V}
\end{equation}
where
\begin{equation}
\begin{split}
D_i W &= W_i + K_i W, \\
K^{ij^*} &= K_{ij^*}^{-1}.
\end{split}
\label{DW}
\end{equation}
Here subscripts denote derivatives with respect to the matter fields, so 
that, for instance, $K_{ij^*}=\partial^2 K/\partial A_i\partial A_j^*$. 
We will be particularly concerned with local minima of \eqref{V} with 
vanishing cosmological constant. Supersymmetry is spontaneously broken 
at such a minimum if and only if $\vev{D_iW} \neq 0$ for some $i$. Under 
these conditions one can show that the gravitino mass is given by
\begin{equation}
m_{3/2}^2 = \vev{e^K |W|^2}.
\label{m3/2}
\end{equation}
The gravitino mass is non-vanishing if and only if supersymmetry is 
spontaneously broken.

\section{Higher-Derivative Bosonic Gravitation}

Einstein bosonic gravitation is specified by the Lagrangian
\begin{equation}
\Lag = - \tfrac{1}{2} \sqrt{-g}\, \R,
\label{Einstein}
\end{equation}
where $g=\det(g_{mn})$ and $\R$ is the spacetime curvature scalar. 
Although $\R$ contains second derivatives of the metric tensor $g_{mn}$, 
the Einstein field equations are nonetheless second-order differential 
equations. Perhaps the simplest modification of Einstein theory is to 
include higher powers of the curvature scalar $\R$ in the Lagrangian. 
Once we include any power of $\R$ greater than unity, the equations of 
motion become fourth-order differential equations. Since more initial 
conditions are required to solve the Cauchy problem, one must conclude 
that the higher-derivative theory has more degrees of freedom than the 
second-order Einstein theory.

A general class of higher-derivative bosonic gravity theories involving 
the scalar curvature only is specified by
\begin{equation}
\Lag = - \tfrac12 \sqrt{-g}\, f(\R),
\label{f(R)}
\end{equation}
where $f$ is an arbitrary real function. These theories do indeed have 
more degrees of freedom than the simple Einstein theory 
\eqref{Einstein}. The dynamical content of Einstein's theory is a single 
massless spin-2 field, the graviton. The higher-derivative theory 
\eqref{f(R)} contains, beside the graviton, one real scalar degree of 
freedom. This degree of freedom can be made explicit using the method of 
Legendre transformations. Here we will follow the discussion of 
\cite{PRD-53-5597}, though such transformations were originally discussed in
\cite{GRG-19-465,PRD-37-1406,CQG-5-L95,CQG-7-557}.
This method reduces the higher-derivative theory to an equivalent second-order 
form. The field equations of the second-order theory are second-order 
differential equations, where the extra degree of freedom is explicitly 
represented by a new field variable. To apply this procedure to the higher-
derivative theory \eqref{f(R)}, we introduce a real scalar field $X$ into the 
Lagrangian
\begin{equation}
\Lag = - \tfrac12 \sqrt{-g}\, \Big\{ f'(X)(\R-X)+f(X) \Big\},
\label{fX}
\end{equation}
where $f'(X)=df/dX$. The equation of motion of the auxiliary field $X$ 
is
\begin{equation}
f''(X)(\R-X)=0.
\end{equation}
Provided that $f''(X)\neq0$, this gives $X=\R$. Substituting back into 
the Lagrangian \eqref{fX} yields the original higher-derivative form 
\eqref{f(R)}. The field redefinition
\begin{equation}
\chi = \ln f'(X),
\end{equation}
puts Lagrangian \eqref{fX} into the form
\begin{equation}
\Lag = - \tfrac12 \sqrt{-g}\, \Big\{ e^\chi \R + f(X(e^\chi)) 
- e^\chi X(e^\chi) \Big\}.
\label{fX2} 
\end{equation}
To remove the non-canonical factor of $e^\chi$ multiplying $\R$, we make 
the conformal transformation
\begin{equation}
\bar g_{mn} = e^\chi g_{mn}.
\end{equation}
Under such a transformation, Lagrangian \eqref{fX2} takes the canonical 
form
\begin{equation}
\label{canonf(R)}
\Lag = \sqrt{-\bar g} \left\{ -\tfrac12 \bR 
       - \tfrac34 \left(\bn\chi\right)^2
       - \tfrac12  e^{-2\chi} \big( f(X(e^\chi)) -  e^\chi X(e^\chi) 
\big) \right\}.
\end{equation}
This Lagrangian describes Einstein gravity coupled to a physically 
propagating real scalar field $\chi$ with a specific potential energy 
dependent on the function $f(\R)$. Note that scalar field $\chi$ has a 
non-ghost-like kinetic energy.

We have studied the vacuum structure of a wide class of 
higher-derivative bosonic gravitational theories in previous papers 
\cite{PRD-53-5583,PRD-53-5597}. We found that they exhibit an interesting vacuum 
structure, but these vacua generically have non-vanishing cosmological constant. 
We extended our investigations to supergravitation in \cite{NPB-471-409}, where 
we studied the vacuum structure of higher-derivative, $D=2$, $N=(1,1)$ 
supergravitation. We found that these theories possess an even richer vacuum 
structure and, more importantly, that non-trivial vacua can occur with vanishing 
cosmological constant. This makes higher-derivative supergravity theories more 
important for particle physics than their bosonic counterparts. In the present 
paper we would like to extend our study to $D=4$, $N=1$ higher-derivative 
supergravity theories.

\section{A Simple Generalization of Einstein Supergravity}

As stated in Section 2, Einstein supergravity is specified by the 
Lagrangian
\begin{equation}
\Lag = - 3 \int d^4\theta E.
\label{SG}
\end{equation}
We would like to generalize this Lagrangian to include higher-derivative 
terms. A simple generalization would be to consider
\begin{equation}
\Lag = - 3 \int d^4\theta E \big( f(R)+f(R^{\dag}) \big),
\label{f}
\end{equation}
where $f$ is a real function of its argument, and $R$ is the chiral 
scalar curvature superfield. Note that the apparently more general
Lagrangian of the form $f+f^{\dag}$, where $f$ is now an analytic function
of $R$, in fact reduces to \eqref{f} by virtue of the indentitiy 
\eqref{id}. We will study action \eqref{f} in this 
section and will consider other generalizations in the following 
section.

What is the dynamical content of Lagrangian \eqref{f}? We need to 
compute the component field Lagrangian to answer this question. We start 
by writing \eqref{f} in chiral superspace as
\begin{align}
\Lag &= \frac34 \int d^2\Theta\E\proj \big( f(R)+f(R^{\dag}) \big) 
+ \text{h.c.} \notag \\
     &= \frac32 \int d^2\Theta \E \proj f(R) + \text{h.c.},
\label{f2}
\end{align}
where we have used the identity \eqref{id}. Since $R$ is chiral, so is 
$f(R)$ and, hence, $\bar{\D}^2 f(R)=0$. Lagrangian \eqref{f2} can then 
be written as
\begin{equation}
\Lag = - 12 \int d^2\Theta \E F(R) + \text{h.c.},
\label{F}
\end{equation}
where $F(R)=Rf(R)$. It is more convenient to carry out the following 
analysis in terms of the function $F(R)$.

The full component-field Lagrangian can be evaluated using the 
$\Theta$-expansions of $\E$ and $R$ given by \eqref{E&R}. However, this 
component Lagrangian is quite involved. For the present purpose, it will 
suffice to consider only the bosonic part of the Lagrangian. In this 
case, the $\Theta$-expansions of $\E$ and $R$ are given by
\begin{equation}
\begin{split}
\E &= \tfrac12 e \big\{1 - \Theta\Theta M^* \big\}, \\
 R &= - \tfrac16 \Big\{ M + \Theta\Theta \big[ - \tfrac12 \R + \tfrac23 
MM^* + \tfrac13 b^m b_m - i \nabla^m b_m \big] \Big\}.
\end{split}
\label{E&R2}
\end{equation}
Consequently we have
\begin{equation}
F(R) = F( -\tfrac16 M ) - \tfrac16  \Theta\Theta F'(-\tfrac16 M) \big[ 
- \tfrac12 \R + \tfrac23 MM^* + \tfrac13 b^mb_m - i \nabla^m b_m \big].
\label{Fbosonic}
\end{equation}
Substituting \eqref{E&R2} and \eqref{Fbosonic} into \eqref{F} and 
performing the $\Theta$ integral yields
\begin{align}
e^{-1} \Lag_{\text{Bosonic}} = &-\tfrac12 (F'+{F'}^*) \R
+ \tfrac13 b^mb_m (F'+{F'}^*) - i \nabla^m b_m (F'-{F'}^*) \notag \\
& + \tfrac23 MM^* (F'+{F'}^*) + 6 (M^* F + M F^*),
\label{LBos}
\end{align}
where, for simplicity, we denote $F(-\tfrac16 M)$ by $F$ and $F(-
\tfrac16 M^*)$ by $F^*$. There are two ways to think about the dynamical 
content of this Lagrangian. The first is to integrate the term involving 
$\nabla^m b_m$ by parts. Doing this clearly makes $b_m$ an auxiliary 
field. After making the integration by parts, varying the Lagrangian 
with respect to $b_m$ yields
\begin{equation}
b_m = - \frac{3i}{2} \frac{\partial_m (F'-{F'}^*)}{F'+{F'}^*}.
\end{equation}
Substituting this expression back into the Lagrangian gives
\begin{align}
e^{-1} \Lag_{\text{Bosonic}} = &-\tfrac12 (F'+{F'}^*) \R + \tfrac34 
(F'+{F'}^*)^{-1} [\partial_m(F'-{F'}^*)]^2 \notag \\
& + \tfrac23 MM^* (F'+{F'}^*) +6 (M^*F+MF^*).
\label{LM1}
\end{align}
We would like to remove the non-canonical coupling of $\R$ to the 
complex field $M$. To do this, we perform the following conformal 
transformation
\begin{equation}
\bar g_{mn} = (F'+{F'}^*) g_{mn}.
\end{equation}
Under such a transformation, Lagrangian \eqref{LM1} takes the form
\begin{align}
{\bar e}^{\, -1} \Lag_{\text{Bosonic}} = & -\tfrac12 \bar\R 
- \tfrac34 (F'+{F'}^*)^{-2} [\partial_m (F'+{F'}^*)]^2 
+ \tfrac34 (F'+{F'}^*)^{-2} [\partial_m (F'-{F'}^*)]^2 \notag \\
& + \tfrac23 MM^* (F'+{F'}^*)^{-1} + 6 (F'+{F'}^*)^{-2}(M^*F+MF^*),
\end{align}
or equivalently
\begin{equation}
{\bar e}^{\, -1} \Lag_{\text{Bosonic}} = -\tfrac12 \bar\R 
- K_{MM^*} \partial^m M \partial_m M^* - V(M,M^*),
\label{L13}
\end{equation}
where $K_{MM^*}=\partial^2 K/\partial M\partial M^*$, $K$ is a real 
function of $M$ and $M^*$ given by
\begin{equation}
K = -3 \ln \big( F'+{F'}^* \big)
\end{equation}
and $V$ is the potential energy given by
\begin{equation}
V(M,M^*) = \frac{12}{(F'+{F'}^*)^2} \big( - \tfrac12 M^* F - 
\tfrac{1}{18}MM^* F' + \text{c.c} \big).  
\label{VM}
\end{equation}
The bosonic structure of the generalized supergravity Lagrangian 
\eqref{F} is now clear. It follows from \eqref{L13} that it describes 
Einstein gravity coupled to a complex scalar field with a specific 
non-linear sigma model kinetic energy and self-interactions. It is 
interesting to note that this simple generalization of supergravity, 
instead of adding higher-derivative gravitational terms such as $\R^2$ 
as discussed in the previous section, promotes the previously auxiliary 
field $M$ to a propagating field, thus adding two real bosonic degrees 
of freedom to the theory. This is the simplest example of a generic 
phenomenon that occurs in higher-derivative supergravity, and will be 
discussed in more detail in the next section. This phenomenon was first 
observed at the linearized level in a component field construction of 
higher-derivative supergravity by Ferrara, Grisaru and van Nieuwenhuizen 
\cite{NPB-138-430}.

Another way to analyze Lagrangian \eqref{LBos} is not to perform the 
integration by parts. Then the Lagrangian does not involve any 
derivatives of $M$ and, hence, $M$ is an auxiliary field. We can then 
eliminate $M$ using its equation of motion. Unfortunately, this equation 
of motion cannot be solved in closed form. Nevertheless, we can see that 
$M$ will be a function of the curvature scalar $\R$ and $\nabla^m b_m$. 
When substituting back into the Lagrangian \eqref{LBos}, we get a 
non-trivial function of the scalar curvature $R$ and $\nabla^m b_m$. The 
non-triviality in $\R$ means this is a higher-derivative Lagrangian for 
the metric tensor. These higher-derivative terms describe, according to 
the analysis of Section~3, the usual graviton plus one real propagating 
scalar degree of freedom. The $\nabla^m b_m$ terms imply that the 
longitudinal mode of $b_m$, once an auxiliary field, is now propagating. 
We again conclude that Lagrangian \eqref{LBos} describes Einstein 
gravity coupled to two real degrees of freedom, in agreement with the 
above analysis.

Since the theory is supersymmetric, there should be superpartners for 
the two new bosonic degrees of freedom. This may seem odd at first, 
because the supergravity multiplet does not contain any auxiliary 
fermionic fields to start propagating along with the bosonic degrees of 
freedom. But having auxiliary fields that acquire kinetic-energy terms 
upon generalizing the Lagrangian is only one way of getting new degrees 
of freedom. Another way is to have higher-derivatives acting on the 
propagating fields. In Einstein supergravity the gravitino kinetic 
energy is given by
\begin{equation}
e^{-1} \Lag_{\text{KE-Fermionic}} = \tfrac12 \varepsilon^{k\ell mn} 
(\bar{\psi}_k \sbar_\ell \tilde{\D}_m \psi_n - \psi_k \sigma_\ell 
\tilde{\D}_m \pb_n),
\end{equation}
as can be seen from \eqref{LE}. The associated equation of motion of 
$\psi_{m\alpha}$ is first-order in derivatives, as it should be for a 
fermionic degree of freedom. Now note that the middle component of 
superfield $R$ contains a first derivative of the gravitino. This can be 
seen from the $\Theta$-expansion of $R$ given in \eqref{E&R}. Hence, 
$R^n$, for any integer $n$ greater than unity, will contain in its 
highest component terms quadratic in the first derivative of the 
gravitino $\psi_{m\alpha}$. Such terms are generically contained in 
Lagrangian \eqref{F}. Thus the component field Lagrangian contains 
higher powers of the first derivative of the gravitino. Therefore, the 
fermionic equation of motion will be second-order, requiring additional  
initial conditions to solve the Cauchy problem. Hence, there will be 
more fermionic degrees of freedom in the theory described by \eqref{F} 
than just a gravitino. In fact, as far as the fermionic degrees of 
freedom are concerned, Lagrangian \eqref{F} always describes a 
higher-derivative theory. These additional fermionic degrees of freedom 
act as the superpartners for the two new bosonic degrees of
freedom discussed above.

It is clear that analyzing the theory specified by \eqref{F} is by no 
means simple. In analogy with the analysis of the bosonic theory 
\eqref{f(R)}, we would like to make the extra degrees of freedom 
explicit. Therefore, we would like a supersymmetric analog of the 
Legendre transform method. In the bosonic case we need to introduce one 
real field to account for the new degree of freedom. Here we have two 
real scalar degrees of freedom, or equivalently one complex scalar, 
along with their fermionic partners. These degrees of freedom can only 
arrange themselves into a chiral supermultiplet. Therefore, we introduce 
a chiral superfield $\Phi$ and try to rewrite Lagrangian \eqref{F} in an 
equivalent second-order form in which the new propagating degrees of 
freedom are contained in $\Phi$. In analogy to the bosonic case 
\eqref{fX}, consider a new Lagrangian
\begin{equation}
\Lag = - 12 \int d^2\Theta \E \Big\{ F'(\Phi) (R-\Phi) + F(\Phi) \Big\}
+ \text{h.c.}
\label{LFPhi}
\end{equation}
The equation of motion of the superfield $\Phi$ is
\begin{equation}
F''(\Phi) ( \Phi - R ) =0.
\end{equation}
Therefore, provided that $F''(\Phi)\neq 0$, this gives
\begin{equation}
\Phi=R.
\label{Phi}
\end{equation}
Substituting \eqref{Phi}  back into \eqref{LFPhi} yields the original 
Lagrangian \eqref{F}. Lagrangian \eqref{LFPhi} can be written as
\begin{equation}
\Lag = - 12 \int d^2\Theta \E \Big\{ RF'(\Phi) - \big[ \Phi F'(\Phi) - 
F(\Phi) \big] \Big\} + \text{h.c.}
\end{equation}
or, equivalently
\begin{equation}
\Lag = \frac34 \int d^2\Theta \E \proj \Big\{ F'(\Phi) + F'(\Phi^{\dag}) \Big\} 
+ 12 \int d^2\Theta \E \Big\{ \Phi F'(\Phi) - F(\Phi) \Big\} + \text{h.c.},
\end{equation}
where we have used the identity \eqref{id} and the fact that $F(R)$ is 
chiral. Comparing this form with \eqref{L3}, we find that this is the 
Lagrangian of Einstein supergravity coupled to a single chiral 
superfield $\Phi$ with a K\"ahler potential given by
\begin{equation}
K(\Phi,\Phi^{\dag}) = -3 \ln \big( F'(\Phi) + F'(\Phi^{\dag}) \big),
\label{KPhi}
\end{equation}
and a superpotential given by
\begin{equation}
W(\Phi) = 6 \big( \Phi F'(\Phi) - F(\Phi) \big).
\label{WPhi}
\end{equation}
In this equivalent formulation of the theory, we can explicitly 
demonstrate the new propagating degrees of freedom. Recall that in 
chiral superspace
\begin{equation}
\Phi = A + \sqrt{2} \Theta^\alpha \chi_\alpha + \Theta^\alpha 
\Theta_\alpha F.
\end{equation}
From the equation of motion of $\Phi$, \eqref{Phi}, we see, using 
\eqref{E&R}, that the new propagating component-field degrees of freedom 
are
\begin{equation}
\begin{split}
A &= - \tfrac16 M, \\
\chi_\alpha &= -\tfrac{\sqrt{2}}{12} \left( \sigma^m \bar\sigma^n
(\tilde\D_m \psi_n - \tilde\D_n\psi_m)
- i \sigma^m \bar\psi_m M + i \psi_m b^m \right)_\alpha.
\end{split}
\label{A&chi}
\end{equation}
The scalar bosonic degree of freedom is the complex scalar field $M$, 
and the fermionic partner, although a complicated combination of 
$\psi_{m\alpha}$, $M$, and $b_m$, is essentially the first derivative of 
the gravitino $\psi_{m\alpha}$. This is in accord with the component 
field discussion presented earlier.

We now turn our attention to the vacuum structure of the theory. To do 
this, we need to evaluate the scalar field potential energy using the 
K\"ahler potential and superpotential given in \eqref{KPhi} and 
\eqref{WPhi} respectively. A straightforward calculation using 
\eqref{V} yields
\begin{equation}
V(A,A^*) = \frac{12}{\big( F'(A)+F'(A^*) \big)^2} \big( 3 A^* F(A) 
- 2 AA^* F'(A) + \text{c.c.} \big).
\label{V2}
\end{equation}
Using \eqref{A&chi}, we see that this potential is exactly the same as 
the potential for the field $M$ in \eqref{VM}, as it must be. The 
kinetic energy for $A$ can be obtained by substituting K\"ahler 
potential \eqref{KPhi} into \eqref{LKEBosonic} which gives 
\begin{equation}
e^{-1} \Lag_{\text{KE-Bosonic}} = - \frac{3}{\big( F'(A)+F'(A^*)\big)^2}
\partial^m A \partial_m A^*.
\end{equation}
Again, noting that $A=-\tfrac16 M$, we see that this is exactly the 
kinetic energy term for $M$ in \eqref{L13}.

An important question is whether or not the potential \eqref{V2} has 
stable vacua, denoted by $\vev{A}$, with zero cosmological constant. By 
a vacuum we will mean a local minimum of the potential. The cases 
$\vev{A}=0$ and $\vev{A}\neq 0$ require separate treatment.

\vspace*{\baselineskip}

\noindent \textsc{Case I: $\vev{A}=0$}

It is clear from \eqref{V2} that $V(0,0)=0$, so the cosmological 
constant vanishes. In order for $\vev{A}=0$ to be a stationary point of 
the potential, the first-order derivatives of the potential evaluated at 
$A=0$, given by 
\begin{equation}
\vev{\frac{\partial V}{\partial A}} =  \frac{9 F(0)}{F'(0)^2},
\end{equation}
should vanish. It follows that we must have 
\begin{equation}
F(0)=0.
\label{F(0)}
\end{equation}
To be a local minimum, the scalar mass matrix should be positive 
definite. In fact, all the second-derivatives of $V$ vanish at $A=0$ 
except
\begin{equation}
\vev{\frac{\partial^2 V}{\partial A \partial A^*}} = 
\frac{6}{F'(0)},
\end{equation}
This should be positive in order to have a local minimum at $A=0$. This 
can be achieved by choosing $F'(0)$ to be positive. Hence, for a wide 
choice of the function $F$ in \eqref{F}, there is a stable vacuum with 
vanishing cosmological constant at $A=0$. Does this vacuum break 
supersymmetry spontaneously? As stated in Section~2, supersymmetry is 
spontaneously broken if and only if 
\begin{align}
D_A W &= \frac{\partial W}{\partial A} + \frac{\partial K}{\partial A} 
W \notag \\
&= 6 A F''(A) - \frac{18 F''(A) (A F'(A) - F(A))}{F'(A)+F'(A^*)},
\end{align}
does not vanish when evaluated at the vacuum. In this case, using 
equation \eqref{F(0)}, we can easily see that $\vev{D_AW}=0$. Therefore, 
supersymmetry is not broken at the $\vev{A}=0$ vacuum. We call the 
$\vev{A}=0$ vacuum solution the trivial vacuum.

\vspace*{\baselineskip}

\pagebreak

\noindent \textsc{Case II: $\vev{A}\neq0$}

Away from $A=0$ we can rewrite the potential \eqref{V2} in the form
\begin{equation}
V(A,A^*) = \frac{12 |A|^2}{\big( F'(A)+F'(A^*) \big)^2} U(A,A^*),
\end{equation}
where
\begin{equation}
U(A,A^*) = 3 \frac{F(A)}{A} - 2 F'(A) + \text{c.c.}.
\label{U}
\end{equation}
In general, a local minimum of $V$ does not need to be a minimum of $U$. 
However, away from $A=0$, $V$ is a strictly positive multiple of $U$.
Thus, it is true that a local minimum of $V$, with the added condition
that $\vev{V}=0$, is a local minimum of $U$ with $\vev{U}=0$.

We see from \eqref{U} that $U$ is a harmonic function. Therefore, it 
cannot have a local minimum, except in the special case when it is a 
constant. We conclude that if $U$ is not a constant function, there 
exists no $\vev{A}\neq 0$ vacuum with zero cosmological constant. Now 
let $U$ be constant. In order to have vanishing cosmological constant 
$U$ must also vanish. It follows from \eqref{U} that $F(\Phi)$ must 
satisfy
\begin{equation}
3 \frac{F(\Phi)}{\Phi} - 2 F'(\Phi) =0,
\end{equation}
which has the unique solution
\begin{equation}
F(\Phi) = c \Phi^{3/2},
\label{Fns}
\end{equation}
where $c$ is a constant which we will set to unity without loss of 
generality.

The choice of $F$ given by \eqref{Fns} is an interesting case. It is a 
higher-derivative theory of supergravity that is equivalent to Einstein 
supergravity coupled to a single superfield with vanishing potential 
energy. The K\"ahler potential and superpotential corresponding to 
\eqref{Fns} are given by
\begin{equation}
\begin{split}
K &= -3 \ln\left\{\tfrac32\Phi^{1/2}+\tfrac32{\Phi^{\dag}}^{1/2}\right\}, \\
W & = 3 \Phi^{3/2}.
\end{split}
\end{equation}
Einstein supergravity coupled to matter is invariant under 
K\"ahler-superWeyl transformations, which can be used to scale the 
superpotential $W$ to unity. Doing this yields a new K\"ahler potential 
$K' = K + \ln W + \ln W^{\dag}$ which we find to be
\begin{equation}
K' = -3 \ln \tfrac{3^{1/3}}{2} \left\{\Phi^{-1/2} + {\Phi^{\dag}}^{-1/2}
\right\},
\end{equation}
as well as the new superpotential
\begin{equation}
W' = 1.
\end{equation}
With the field redefinition
\begin{equation}
S = \frac{3^{1/3}}{2} \Phi^{-1/2},
\end{equation}
the K\"ahler potential and superpotential take the standard form of 
no-scale supergravity \cite{PLB-133-61}
\begin{equation}
\begin{split}
K' &= -3 \ln \big( S + S^{\dag} \big), \\
W' &= 1. 
\end{split}
\end{equation}
Since the scalar field potential energy of this theory vanishes, any 
$\vev{S} \neq 0$ value is a vacuum solution with zero cosmological 
constant. It follows from \eqref{DW} that
\begin{align}
D_S W' &= \frac{\partial K'}{\partial S} \notag \\
&= -\frac{3}{S+S^{\dag}}.
\end{align}
Therefore, $\vev{D_S W'}\neq 0$ for any $\vev{S}\neq 0$ and, hence, 
supersymmetry is always spontaneously broken. It is amusing to note that 
the simple no-scale theory is, in fact, completely equivalent to a 
higher-derivative theory of pure supergravitation specified by 
Lagrangian \eqref{F} with function $F$ given in \eqref{Fns}.

\section{A Generic Higher-Derivative Supergravity}

In Section~4, we studied a class of supergravity theories where the 
Lagrangian is the sum of an arbitrary function of the chiral superfield 
$R$ and its hermitian conjugate. The vacuum structure of this class of 
theories was found to be very restricted. If we insist on having zero 
cosmological constant, we are generically forced to the trivial vacuum 
at $A=0$. This vacuum does not spontaneously break supersymmetry.

In this section, we consider a second, more general, class of 
supergravity theories in which the Lagrangian is given by
\begin{equation}
\Lag = -3 \int d^4\theta E f(R,R^{\dag}),
\label{f3}
\end{equation}
where $f$ is a real function. Of course, this class will include the 
previous one as a special case. In the following, we will restrict our 
discussion to functions $f(R,R^{\dag})$ that cannot be split into 
$f(R)+f(R^{\dag})$. Most of the following analysis will break down for the 
$f(R)+f(R^{\dag})$ case, and we will make occasional reference to where 
such breakdown occurs.

What is the dynamical content of Lagrangian \eqref{f3}? We need to 
compute the component field Lagrangian to answer this question. We start 
by writing \eqref{f3} in the chiral superspace as
\begin{equation}
\Lag = \frac34 \int d^2\Theta \E \proj f(R,R^{\dag}) + \text{h.c.}
\label{f11}
\end{equation}
The full component field Lagrangian can be evaluated using the 
$\Theta$-expansions of $\E$ and $R$ given in \eqref{E&R}. However, this 
component Lagrangian is very complicated. For the present purpose, it 
suffices to consider only the bosonic part of the Lagrangian. In this 
case, we can use the expressions for $\E$ and $R$ given in \eqref{E&R2}. 
Substituting these into \eqref{f11} and performing the 
$\Theta$-integral, we find that
\begin{align}
e^{-1} \Lag_{\text{Bosonic}} = & - \tfrac12 
(f+Mf_M+M^*f_{M^*}-4MM^*f_{MM^*}- 2 b^mb_m f_{MM^*}) \R
- \tfrac{3}{4} f_{MM^*} \R^2 \notag \\
& + 3 f_{MM^*} \partial^m M \partial_m M^* - 3 f_{MM^*} 
(\nabla^m b_m)^2 +i (f_M \partial^mM - f_{M^*} \partial^m M^*) b_m 
\notag \\
& -i (f_M M - f_{M^*} M^* ) \nabla^m b_m - \tfrac13 MM^*
[f - 2(M f_M + M^* f_{M^*}) + 4 MM^* f_{MM^*}] \notag \\
& + \tfrac13 b^m b_m (f + M f_M + M^* f_{M^*} - 4 MM^* f_{MM^*}
- b^m b_m f_{MM^*}),
\end{align}
where $f=f(-\tfrac16 M,-\tfrac16 M^*)$, $f_M=\partial f/\partial M$, and 
$f_{M^*}=\partial f/\partial M^*$. The first notable feature of this 
Lagrangian is the appearance of an $\R^2$ term in addition to $\R$. 
Hence, this is a higher-derivative theory of gravity. Note that there 
are no higher powers of $\R$. Therefore, this is a theory of quadratic 
gravitation only. Recall that $\R+\R^2$ gravity describes a graviton 
plus one real scalar degree of freedom. The second feature is that there 
are no auxiliary fields. Both $M$ and $b_m$ have kinetic energy terms. 
The field $M$ has the normal kinetic energy for a complex scalar field 
with a sigma-model factor in front. The field $M$ represents 
one complex scalar 
degree of freedom or two real degrees of freedom. The vector field $b_m$ 
has a kinetic energy term for its longitudinal mode. This is one more 
propagating degree of freedom. Note that the coefficient of $\R^2$ and 
the kinetic energy terms of $M$ and $b_m$ are proportional to 
$f_{MM^*}$. This will vanish if $f(R)$ splits into $f(R)+f(R^{\dag})$, 
a case which we 
exclude from the discussion as mentioned before. To conclude, the 
bosonic part of Lagrangian \eqref{f3} describes the propagation of four 
real degrees of freedom in addition to the usual graviton of Einstein 
theory. What about the fermionic degrees of freedom? Since the theory is 
supersymmetric, there must be extra fermionic degrees of freedom that 
propagate along with the new bosonic degrees of freedom. These 
fermionic degrees of freedom should be associated with 
higher-derivatives acting on $\psi_{m\alpha}$, as it is the only 
fermionic field in the supergravity multiplet. An analysis of the 
fermionic part of the Lagrangian shows that the gravitino field equation 
is, indeed, not first-order but third-order. We will exhibit these new 
fermionic degrees of freedom explicitly below.

In analogy with the analysis in the previous section, we would like to 
make the extra degrees of freedom explicit. Therefore, we would like a 
supersymmetric analog of the Legendre transform method. Here we have 
four real bosonic degrees of freedom, or equivalently two complex 
scalars, along with their fermionic partners. These degrees of freedom 
can only arrange themselves into two chiral supermultiplets. Therefore, 
we introduce two chiral superfields $\Phi$ and $\Lambda $ and try to 
rewrite the Lagrangian \eqref{f11} in an equivalent second-order form in 
which the new propagating degrees of freedom are contained in $\Phi$ and 
$\Lambda$. This equivalence was first established in \cite{PLB-190-86} 
using the 
compensator formalism of supergravity. It was also discussed in 
supergravity theories with Chern-Simons terms, such as those associated 
with low energy superstring effective Lagrangians 
\cite{NPB-138-430,PLB-254-132}. Consider a new Lagrangian
\begin{align}
\Lag &= \frac34 \int d^2\Theta \E \proj f(\Phi,\Phi^{\dag}) 
+ 6 \int d^2\Theta \E \Lambda (\Phi-R)  + \text{h.c.} \notag \\
&= \frac34 \int d^2\Theta \E \proj 
\big( f(\Phi,\Phi^{\dag})+\Lambda+\Lambda^{\dag} \big)
+ 6 \int d^2\Theta \E \Phi \Lambda + \text{h.c.}
\label{f4}
\end{align}
The equation of motion of $\Lambda$ can be obtained by varying 
Lagrangian \eqref{f4} with respect to $\Lambda$. From the first line of 
\eqref{f4} we have
\begin{equation}
\delta_{\Lambda}{\mathscript{L}} = 6 \int{d^{2}\Theta {\mathscript{E}} 
\left(\Phi-R\right)\delta\Lambda}.
\end{equation}
Setting $\delta_\Lambda\Lag$ to zero gives the equation of motion
\begin{equation}
\Phi=R.
\label{eom1}
\end{equation}
Substituting this back into \eqref{f4} yields the original Lagrangian 
\eqref{f11}. For completeness, we display the $\Phi$ equation of motion 
which is given by
\begin{equation}
\Lambda= -\tfrac18 \left(\bar{\mathscript{D}}^{2}-8R\right)\frac{\partial
f}{\partial \Phi}.
\label{eom2}
\end{equation}

Now let us compare Lagrangian \eqref{f4} with the standard form of the 
Lagrangian for chiral matter coupled to supergravity in chiral 
superspace \eqref{SGMC}. We clearly see that Lagrangian \eqref{f4} 
describes Einstein supergravity coupled to two chiral superfields $\Phi$ 
and $\Lambda$ with the K\"ahler potential and superpotential
\begin{equation}
\begin{split}
K &= -3\ln\big(f(\Phi,\Phi^{{\dag}ger})+\Lambda +
\Lambda^{{\dag}ger}\big), \\
W &= 6\Phi\Lambda,
\end{split}
\label{K&W}
\end{equation}
respectively. Although the case in which the function $f(\Phi,\Phi^{\dag})$ 
splits into $f(\Phi)+f(\Phi^{\dag})$ is excluded from the present discussion, 
we would like to 
point out that in such a case \eqref{K&W} is still valid. However, one
can also perform the field redefinetion
\begin{equation}
\Phi \rightarrow f(\Phi)+\Lambda,
\end{equation}
to render the K\"ahler potential independent of $\Lambda$. 
Thus, the superfield $\Lambda$ becomes auxiliary 
with an algebraic equation of motion. One can use this equation of 
motion to eliminate $\Lambda$ from the Lagrangian and end up with a theory
of Einstein supergravity coupled to a single propagating chiral superfield.
This is in agreement with the discusion of the previous section.

The extra degrees of freedom in the higher-derivative theory \eqref{f3} 
are now manifest in the lowest and fermionic components of the 
superfields $\Phi$ and $\Lambda$
\begin{equation}
\begin{split}
\Phi &= A + \sqrt{2} \Theta^\alpha \chi_\alpha + \Theta^\alpha 
\Theta_\alpha F, \\
\Lambda &= B + \sqrt{2} \Theta^\alpha \xi_\alpha + \Theta^\alpha 
\Theta_\alpha G.
\end{split}
\end{equation}
Using the equations of motion \eqref{eom1} and \eqref{eom2}, we find 
that to lowest order the new propagating degrees of freedom are given by
\begin{equation}
\begin{split}
A &= - \tfrac16 M, \\
\chi_\alpha &= -\tfrac{\sqrt{2}}{12} 
\big(\sigma^m \bar\sigma^n (\partial_m \psi_n
- \partial_m \psi_n )\big)_\alpha,
\end{split}
\end{equation}
and
\begin{equation}
\begin{split}
B &= - \tfrac16 f_R(0,0) M - \tfrac{1}{12} f_{RR^{\dag}}(0,0)
( - \tfrac12 \R + i \partial^m b_m ), \\
\xi_\alpha &= -\tfrac{\sqrt{2}}{12} f_R(0,0) 
\ud\sigma{m}{\alpha\dot{\alpha}} 
{\bar\sigma}^{n\dot\alpha\beta} 
(\partial_m \psi_{n\beta} - \partial_n \psi_{m\beta} )
- \tfrac{i\sqrt{2}}{24} f_{RR^{\dag}}(0,0) \ud\sigma{m}{\alpha\dot\alpha}
\Big( \partial^n \partial _m \du{\bar\psi}{n}{\dot\alpha} 
- \partial^n \partial_n 
\du{\bar\psi}{m}{\dot\alpha} \Big).
\end{split}
\end{equation}
As in the previous section, the scalar and fermionic component fields in 
$\Phi$ are the $M$ field and, essentially, the first derivative of the 
gravitino $\psi_{m\alpha}$ respectively. On the other hand, the scalar 
field associated with $\Lambda$ is constructed from $M$, $\R$ and 
$\nabla^m b_m$, whereas the fermionic field is, essentially, the second 
derivative of the gravitino.

To study the vacuum structure of the theory we need to evaluate the 
bosonic part of the component field Lagrangian corresponding to 
\eqref{f11}. The bosonic part of the Lagrangian is given in 
\eqref{LBosonic} where $K$ and $W$ are given in \eqref{K&W}. The 
kinetic energy for the bosonic fields $A$ and $B$ is obtained by 
substituting \eqref{K&W} into \eqref{LKEBosonic}. The result is
\begin{equation}
e^{-1} \Lag_{\text{KE-Bosonic}} = - 3 ( f + B + B^* )^{-2} \Big\{ 
- \left( f + B + B^* \right) f_{AA^*} \partial^m A
\partial_m A^* + | f_A \partial_m A + \partial_m B|^2 \Big\}.
\label{KEB}
\end{equation}
We see that for non-ghost-like propagation of the fields $A$ and $B$
we require that at the vacuum 
\begin{equation}
\vev{(f+B+B^*)f_{AA^*}} < 0.
\label{ng}
\end{equation}
We can also substitute \eqref{K&W} into \eqref{V} to give the potential 
energy
\begin{equation}
V = 12 \left(f(A,\Ad)+B+\Bd\right)^{-2} U(A,B),
\label{V3}
\end{equation}
where $U$ is given by
\begin{equation}
U = \left|A\right|^{2}\Big( f-2\left( f_{A}A+f_{A^*}A^* \right)
+ 4 f_{AA^*} |A|^2 \Big)
- f^{-1}_{A\Ad} \Big|B-f_{A}A+2f_{AA^*}|A|^2\Big|^{2}.
\label{UAB}
\end{equation}

An important question is whether or not potential \eqref{V3} has stable 
vacua, generically denoted by $\vev{A}$ and $\vev{B}$, with zero 
cosmological constant that break supersymmetry spontaneously. In 
general, a local minimum of $V$ need not be a local minimum of $U$ and 
vice versa. However, $V$ is a strictly positive multiple of $U$. 
As in Section~4, it is simple to argue that
$\vev{A}$, $\vev{B}$ is a local minimum of $V$ with zero 
cosmological constant if and only if it is a local minimum of $U$ at 
which $U$ vanishes. The function $U$ has a simpler structure than $V$ 
and it is possible to make some general statements about its zero 
cosmological constant minima. First note that the second term in $U$ is 
unbounded from below unless 
\begin{equation}
\vev{f_{AA^*}^{-1}} < 0.
\label{s}
\end{equation}
If this condition is satisfied, the condition for non-ghost-like
propagation \eqref{ng} simplifies to
\begin{equation}
\vev{f+B+B^*} > 0.
\label{ng-simple}
\end{equation}
Assuming that $\vev{f_{AA^*}^{-1}}$ is negative definite, since $B$ 
enters only the second term of \eqref{UAB}, minimizing $U$ with respect 
to $B$ gives
\begin{equation}
\vev{B} = \vev{f_A A - 2 f_{AA^*} A A^*},
\label{ss}
\end{equation}
at which point the second term vanishes. It follows that $\vev{A}$ must 
be a local minimum of the first term on the right hand side of 
\eqref{UAB} with zero cosmological constant. Thus, the problem of 
finding a local minimum of $V$ with vanishing cosmological constant 
and where the $A$ and $B$ fields have non-ghost-like propagation
is reduced to finding a local minimum $\vev{A}$ of
\begin{equation}
U_0 = |A|^2 \Big( f - 2 [f_A A + f_{A^*} A^*] + 4 f_{AA^*} |A|^2 \Big),
\label{U0}
\end{equation}
at which $U_0=0$ and which satisfies \eqref{s} and \eqref{ng-simple}. 
The value of $\vev{B}$ then follows from \eqref{ss}.

The above discussion enables us to construct a wide class of Lagrangians 
which have a rich vacuum structure. We demonstrate this by considering a 
concrete example. Since we will be interested in the different 
energy-scales in this model, we restore the gravitational coupling 
constant $\kappa=M_P^{-1}$ where $M_P$ is the Planck mass. Consider
\begin{equation}
f(R,R^{\dag}) = 1 - 2 \frac{RR^{\dag}}{m^2} + \frac19 
\frac{(RR^{\dag})^2}{m^4},
\label{fex}
\end{equation}
where $m$ is a coupling parameter with mass dimension one. This coupling 
parameter need not be related to the Planck mass. We would also like all 
the fields to have canonical dimensions. In the above discussion, $\Phi$ 
has mass-dimension one and $\Lambda$ is dimensionless. In order to give 
$\Lambda$ dimension one, we scale it by $M_P$. Furthermore, it is 
convenient to write the K\"ahler potential in terms of $M_P$ only, 
relegating mass $m$ to the superpotential. This be achieved if 
we scale $\Phi$ by $M_P/m$. The K\"ahler and superpotential associated 
with \eqref{fex} are then given by
\begin{equation}
\begin{split}
K &= -3 M_P^2 \ln \left\{ 1 - 2 \frac{\Phi\Phi^{\dag}}{M_P^2} 
+ \frac19 \frac{(\Phi\Phi^{\dag})^{2}}{M_P^4} 
+ \frac{\Lambda + \Lambda^{\dag}}{M_P} \right\}, \\
W &= 6 m \Phi \Lambda.
\end{split}
\label{K&W3}
\end{equation}
Using \eqref{V3} and \eqref{fex} the potential energy becomes
\begin{multline}
V = 12 m^2 M_P^2 \left(1 - 2 \frac{|A|^{2}}{M_P^2} + 
\frac{1}{9}\frac{|A|^4}{M_P^4} + \frac{B + \Bd}{M_P} \right)^{-2}
\left[
\frac{|A|^2}{M_P^2} \left(\frac{|A|^2}{M_P^2}-1\right)^2 \right. \\
\left. + \frac92 \left(9-2\frac{|A|^2}{M_P^2}\right)^{-1} 
\left| \frac{B}{M_P} - \frac23 \frac{|A|^2}{M_P^2} 
\left(3-\frac{|A|^2}{M_P^2}\right) \right|^2 \right],
\label{Vex}
\end{multline}
and the associated function $U_0$ is given by
\begin{equation}
U_0 = \frac{|A|^2}{M_P^2} \left( \frac{|A|^2}{M_P^2} - 1 \right)^2.
\label{c}
\end{equation}
This function is non-negative and clearly has two local minima at which 
$U_0$ vanishes. The first is at $\vev{A}=0$. Now
\begin{equation}
f_{AA^*} = - \frac{2}{M_P^2} \left( 1 - \frac29 \frac{|A|^2}{M_P^2} 
\right).
\label{cc}
\end{equation}
Therefore, at $\vev{A}=0$, $\vev{f_{AA^*}^{-1}}<0$ as required. It 
follows from \eqref{ss} that $\vev{B}=0$. We conclude that the above 
potential has a local minimum at
\begin{equation}
\vev{A} = \vev{B} = 0.
\label{trivial}
\end{equation}
with vanishing cosmological constant. This minimum is clearly visible at 
the center of the potential energy plotted in Figure 1. In order for 
\eqref{trivial} to be a physically acceptable vacuum, the kinetic-energy 
terms for the scalar fields must have the correct sign when evaluated at 
this point, otherwise fields $A$ and $B$ would be ghost-like. 
We argued above that the kinetic-energy terms have the correct sign if
and only if $\vev{f+B+B^*}>0$. In fact
\begin{equation}
\vev{f+B+B^*}=1>0
\end{equation}
as required. Explicitly, the $A$ 
and $B$ field kinetic-energy terms are given in \eqref{KEB}. When 
evaluated around vacuum \eqref{trivial} to quadratic order, they become
\begin{equation}
e^{-1} \Lag_{\text{KE-Bosonic}} = - 6 \partial^m A \partial_m A^* 
- 3 \partial^m B \partial_m B^*,
\end{equation}
which is clearly non-ghost-like. Is supersymmetry broken by this vacuum? 
Using \eqref{DW} and \eqref{K&W3}, we find that
\begin{equation}
\begin{split}
D_A W &= 6 m B \left\{ 1 + 6 \frac{|A|^2}{M_P^2} 
\frac{\left(1-\frac19 \frac{|A|^2}{M_P^2}
\right)}{\left( 1 - 2 \frac{|A|^2}{M_P^2} 
+ \frac19 \frac{|A|^4}{M_P^4} + \frac{B+B^*}{M_P} \right)} \right\}, \\
D_B W &= 6 m A \left\{ 1 - 3 \frac{B}{M_P} \frac{1}{ \left(
1 - 2 \frac{|A|^2}{M_P^2} 
+ \frac19 \frac{|A|^4}{M_P^4} + \frac{B+B^*}{M_P} \right)} \right\}.
\end{split}
\label{DWex}
\end{equation}
It follows that for $\vev{A}=\vev{B}=0$ both $\vev{D_A W}$ and $\vev{D_B 
W}$ vanish. Hence, supersymmetry is not spontaneously broken. We call 
vacuum \eqref{trivial} the trivial vacuum.

\begin{figure}[ht]
   \vspace*{0.00in}
   \centerline{\psfig{figure=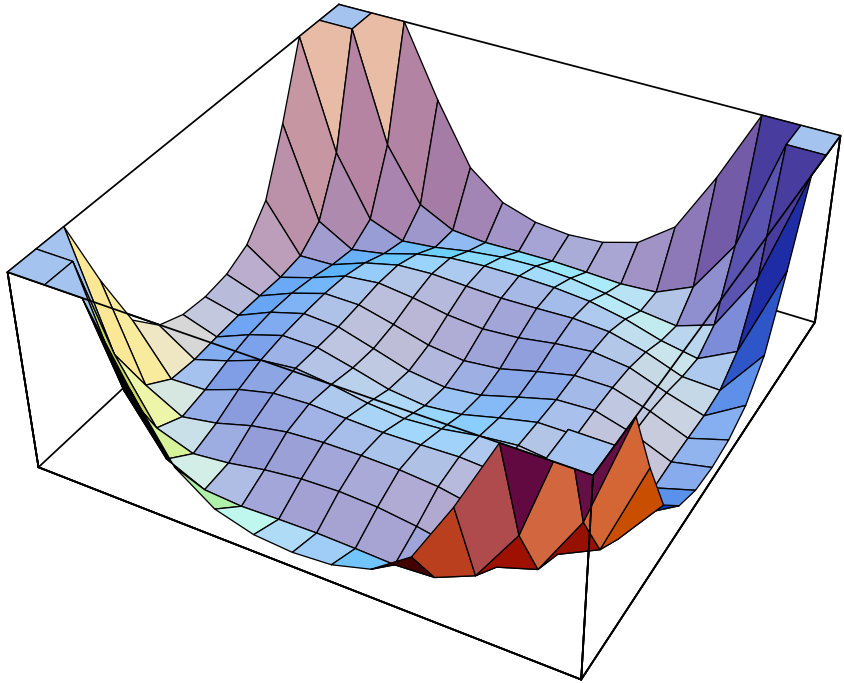,height=4in}}
   \vspace*{0.00in}
   \caption{The potential energy as a function of $A$ at 
            $B=f_AA-2f_{AA^*}AA^*$.}
   \label{fig}
\end{figure}

Are there any other minima of potential \eqref{Vex}? As can be seen from 
Figure 1, there is a second local minimum of $U_0$ given in \eqref{c} at 
which $U_0$ vanishes. It occurs for $|\vev{A}|^2=M_P^2$ or $\vev{A}=M_P 
e^{i\theta}$ for arbitrary real phase $\theta$. Note that this 
degeneracy in the vacuum is a direct consequence of the symmetry of the 
function $f$ in \eqref{fex}, and hence the Lagrangian, under the 
transformation $R \rightarrow e^{i\theta} R$. It follows from \eqref{cc} 
that at this minimum $\vev{f_{AA^*}^{-1}}<0$ as required. Equation 
\eqref{ss} then gives $\vev{B} = \frac43 M_P$. Therefore, the potential 
energy has a ring of local minima at
\begin{equation}
\begin{split}
\vev{A} &= M_P e^{i\theta}, \\
\vev{B} &= \tfrac43 M_P,
\end{split}
\label{vac}
\end{equation}
with vanishing cosmological constant. These are visible as the circular 
set of minima in Figure 1. To check that there are no ghosts in the 
theory around vacuum \eqref{vac}, we need to show that condition 
\eqref{ng-simple} is satisfied. We find that
\begin{equation}
\vev{f+B+B^*} = \tfrac{16}{9} > 0
\end{equation}
as required. Explicitly, expanding $A=\vev{A}+a$ and 
$B=\vev{B}+b$, it follows from \eqref{KEB} and \eqref{vac} that to 
quadratic order 
\begin{equation}
e^{-1} \Lag_{\text{KE-}AB} = - \tfrac{21}{8} \partial^m a 
\partial_m a^* - \partial^m c \partial_m c^*,
\label{KEvac}
\end{equation}
where $c= \sqrt{3} ( \frac{9b}{16} - e^{-i\theta} a )$, which is 
clearly non-ghostlike. The masses of the four real scalars are easily 
evaluated at this minimum, but the exact values are complicated and 
unenlightening. Suffice it to say that, in addition to the one zero-mass 
field in the direction of the circular set of minima, the three 
remaining masses are numerically different but of order $m$. Is 
supersymmetry broken at the vacua \eqref{vac}? Using \eqref{DWex}, we 
find that the K\"ahler covariant derivatives evaluated at \eqref{vac} 
are
\begin{equation}
\begin{split}
\vev{D_A W} &= 32  m M_P, \\
\vev{D_B W} &= -\tfrac{15}{2} e^{i\theta} m M_P.
\end{split}
\end{equation}
It follows that supersymmetry is indeed spontaneously broken at these 
vacua \eqref{vac}. Substituting \eqref{K&W3} into \eqref{m3/2} gives the 
gravitino mass 
\begin{equation}
m_{3/2} = \frac{27}{8} m,
\end{equation}
which is of the same order as the non-vanishing scalar masses. Note that 
the supersymmetry breaking scale is set by $\vev{D_A W}$ and $\vev{D_B 
W}$, which are of order $m M_P$. It is conventional to denote this scale 
by $\mu=\sqrt{mM_P}$. The gravitino mass can then be written in the 
familiar form
\begin{equation}
m_{3/2} = \frac{27}{8} \frac{\mu^2}{M_P}.
\end{equation}
With this definition, the non-vanishing scalar field masses at the 
non-trivial vacuum are of order $\mu^2/M_P$.

The preceding discussion is exact and valid at any energy scale. Indeed, 
as we have shown above, it is necessary to know the full theory in order 
to determine the vacuum structure. Once one has found the non-trivial 
supersymmetry breaking vacuum \eqref{vac}, however, it is of interest to 
consider small fluctuations around this vacuum, and to determine the 
effective theory for these fluctuations at energy scales much smaller 
than the Planck mass $M_P$. We now proceed to do this. Above we did not 
specify the value of the supersymmetry breaking scale $\mu$ relative to 
$M_P$. For the remainder of this discussion we will assume that $\mu\ll 
M_P$. For example, if we would like the gravitino mass to be of the 
order of the electroweak scale, that is $m_{3/2} \simeq 10^2 \text{ 
GeV}$, then we must choose $\mu \simeq 10^{11} \text{ GeV}$, eight 
orders of magnitude smaller than $M_P$. Expanding around the non-trivial 
vacuum \eqref{vac} with $\theta=0$, we define 
\begin{equation}
\begin{split}
\Phi &= M_P + \phi, \\
\Lambda &= \tfrac43 M_P + \lambda, \\
\end{split}
\label{PLp}
\end{equation}
where $\phi$ and $\lambda$ are the fluctuations of $\Phi$ and 
$\Lambda$ around their vacuum expectation values, $M_P$ and $\tfrac43 
M_P$ respectively. We will henceforth consider $\phi$ and $\lambda$ to 
be of the order of $\mu$ or smaller. The exact Lagrangian is given by 
expression \eqref{SG+Matter} with K\"ahler potential and superpotential 
\eqref{K&W3}. At the scale $\mu$, since $\mu/M_P\ll 1$, the 
gravitational effects become negligible, and the supergravity Lagrangian 
\eqref{SGMC} simplifies to the flat superspace Lagrangian given by
\begin{equation}
\Lag = \int d^2\theta d^2\bar\theta K + \left\{ \int d^2\theta W 
+ \text{h.c} \right\}.
\label{Lflat}
\end{equation}
However, in this regime we must also neglect terms in $K$ and $W$ 
suppressed by powers of $\mu/M_P$. That is the flat superspace 
Lagrangian \eqref{Lflat} in fact reduces to
\begin{equation}
\Lag = \int d^2\theta d^2\bar\theta K' + \left\{ \int d^2\theta
W' + \text{h.c.} \right\},
\label{Lf}
\end{equation}
where $K'$ and $W'$ are obtained by substituting \eqref{PLp} into 
\eqref{K&W3} and keeping only the first non-trivial terms in $\mu/M_P$. 
Note that any terms in the K\"ahler potential which are constant or 
purely chiral plus antichiral, such as $\Phi'+{\Phi'}^{\dag}$, do not 
contribute to Lagrangian \eqref{Lf}. A constant term in the 
superpotential also does not contribute. In this limit, we find that
\begin{equation}
\begin{split}
K' &= \tfrac{45}{8} \phi \phi^{\dag} - \tfrac{27}{16} 
(\phi \lambda^{\dag} + \lambda \phi^{\dag} ) 
+ \tfrac{243}{256} \lambda \lambda^{\dag}, \\
W' &= 6 \mu^2 ( \tfrac43 \phi + \lambda ).
\label{K&W'}
\end{split}
\end{equation}
We can diagonalize the K\"ahler potential by making the following field 
redefinitions
\begin{equation}
\begin{split}
\Pi &= \sqrt{\tfrac{21}{8}} \phi, \\
\Sigma &= \sqrt{3} (\tfrac{9}{16} \lambda - \phi).
\end{split}
\end{equation}
In terms of superfields $\Pi$ and $\Sigma$, K\"ahler potential and 
superpotential \eqref{K&W'} become
\begin{equation}
\begin{split}
K' &= \Pi \Pi^{\dag} + \Sigma \Sigma^{\dag}, \\
W' &= \tilde\mu^2 \left( \sqrt{\tfrac72} \Pi + \Sigma \right),
\end{split}
\label{K&W'2}
\end{equation}
where $\tilde\mu^2 = \tfrac{32\sqrt{3}}{9} \mu^2$. The scalar field 
potential energy corresponding to \eqref{K&W'2} is simply a constant 
given by
\begin{equation}
V = \tfrac92 \tilde\mu^4.
\label{Vhd}
\end{equation}
Therefore, below the Planck scale, the theory behaves like a simple 
version of the \linebreak
O'Raifeartaigh model \cite{NPB-96-331} with two chiral superfields, $\Pi$ and 
$\Sigma$. The positive definite value of the scalar potential signals 
supersymmetry breaking at scale $\mu$. Note that the K\"ahler potential 
and superpotential \eqref{K&W'2} look like a simple extension of the 
Polonyi model with two superfields. For comparison, let us consider the 
original Polonyi model for supersymmetry breaking \cite{PP-KFKI-93}. The 
Polonyi model has one chiral superfield $Z$ with K\"ahler potential and 
superpotential given by
\begin{equation}
\begin{split}
K &= ZZ^{\dag}, \\
W &= \mu^2 (Z+\beta),
\end{split}
\label{K&WP}
\end{equation}
where $\beta$ is a constant that is chosen to ensure vanishing 
cosmological constant. The exact theory is found to have a unique vacuum 
at
\begin{equation}
\vev{Z} = (\sqrt3 -1) M_P,
\end{equation}
which has vanishing cosmological constant if $\beta$ is chosen to be 
$\beta = (2-\sqrt3)M_P$. Supersymmetry is spontaneously broken with 
strength $\mu$ at this vacuum since
\begin{equation}
\vev{D_Z W} = \sqrt3 \mu^2.
\end{equation}
Expanding $Z = \vev{Z} + z$, and performing the same analysis as we did 
above, we find that the effective flat superspace K\"ahler potential and 
superpotential are given by
\begin{equation}
\begin{split}
K' &= z z^{\dag}, \\
W' &= \mu^2 z.
\end{split}
\label{K&WP'}
\end{equation}
The associated scalar field potential energy is
\begin{equation}
V = \mu^4.
\label{VP}
\end{equation}
Comparing \eqref{K&W'2} and \eqref{Vhd} with \eqref{K&WP'} and 
\eqref{VP} respectively, we see that our higher-derivative 
supergravitation model is, at low energy, simply a two superfield 
Polonyi model, with no essential differences. They only differ at the 
Planck scale. The Polonyi model can be thought of as the most trivial 
extension of the flat superspace O'Raifeartaigh model. In the Polonyi 
extension, one does not change the low energy superpotential or K\"ahler 
potential at the Planck scale, except for the addition of the constant 
term $\beta$ which is required to set the cosmological constant of the 
non-trivial vacuum to zero. However, this extension is by no means 
unique. As one goes up in energy, one could start seeing generic 
modifications to both the superpotential and the K\"ahler potential 
consisting of new terms that are suppressed by powers of the Planck 
mass. The theory constructed in this section is an explicit example of 
such a phenomenon.

We end this section by noting that there is a class of general
higher-derivative supergravity theories which have a no-scale
structure, with flat directions in the scalar potential. In the
previous section, when $f(R,R^{\dag})$ could be decomposed into
$f(R)+f(R^{\dag})$, we found that choosing $F(R)=Rf(R)=cR^{-3/2}$
made the scalar potential identically zero. In fact, the transformed
theory, in terms of the new chiral field $\Phi$, was identical to
pure no-scale supergravity. In the general case, we see that we cannot
chose $f(R,R^{\dag})$ such that the whole scalar potential is zero
because the second, quadratic, term in $U$ \eqref{UAB} cannot be made to
vanish. However, choosing $f(R,R^{\dag})$ can introduce flat directions
into the potential, giving a continuous set of degenerate vacua with
zero cosmological constant. The maximum possible degeneracy is
acheived by setting the first term in the potential, $U_0$ \eqref{U0},
to zero. The general form of $f(R,R^{\dag})$ with this property is
\begin{equation}
   f\left(R,R^{\dag}\right) = \sqrt{RR^{\dag}} \left[
        h(R) + \left(h(R)\right)^{\dag} \right],
\end{equation}
where $h(R)$ is an arbitrary complex function of $R$. The K\"ahler
potential and superpotential for the transformed theory are then given
by 
\begin{equation}
\begin{aligned}
   K &= -3\ln\left\{\sqrt{\Phi\Phi^{\dag}}\left[ h(\Phi) 
               + \left(h(\Phi)\right)^{\dag} \right]
           + \Lambda + \Lambda^{\dag} \right], \\
   W &= 6\Phi\Lambda.
\end{aligned}
\end{equation}
and the scalar potential becomes 
\begin{equation}
   V = -\frac{48|A|}{\left[|A|(h+h^*)+B+B^*\right]^2
           \left[h+Ah'+h^*+A^*{h'}^*\right]}
           \left|B + \left|A\right|A^*{h'}^*\right|^2, 
\label{noscaleV}
\end{equation}
where $h'=dh/dA$. Considered as a function of $B$, the potential is
clearly only bounded from below if the prefactor in \eqref{noscaleV} is
positive. As before this gives the condition that at the vacuum 
\begin{equation}
   \vev{f_{AA^*}^{-1}} = \vev{|A|\left(h+Ah'+h^*+A^*{h'}^*\right)} 
      < 0.
\label{noscalecond}
\end{equation}
As before to ensure that the $A$ and $B$
fields have non-ghost-like propagation we require $\vev{f+B+B^*}>0$. 
It is clear that setting
$B=-|A|A^*{h'}^*$ sets the potential to zero, independent of the value
of $A$. Thus the potential has flat directions, with a continuous set
of vacua, stable provided \eqref{noscalecond} is satisfied, and given
by 
\begin{equation}
\begin{aligned}
   \vev{A} &= A_0, \\
   \vev{B} &= - \left|A_0\right| A_0^* \left[h'(A_0)\right]^*.
\end{aligned}
\end{equation}
As in the pure no-scale model, all these vacua break
supersymmetry, with zero cosmological constant. The value of the
gravitino mass depends on the vaccum in question and is given by 
\begin{equation}
   m_{3/2} = \frac{\left|\vev{A}\right|^3\left|\vev{h'}\right|^2}
                {\vev{h-Ah'+h^*-A^*{h'}^*}}.
\end{equation}
The important point here, as in all no-scale models, is that the flat
directions in the scalar potential imply that the vacuum is
undetermined. As a result the value of the gravitino mass is also
undetermined at tree level. 

\section{Conclusion}

We have shown in this paper that higher-derivative $N=1$ supergravity in 
four dimensions has non-trivial vacua with vanishing cosmological 
constant that spontaneously break supersymmetry. This result opens the 
possibility of a new approach to supersymmetry breaking in 
phenomenological supergravity theories and, perhaps, in superstring 
theories. In this new approach, supersymmetry is spontaneously broken by 
a non-trivial vacuum of the new degrees of freedom associated with 
higher-derivative supergravitation, and does not need a hidden sector or 
gaugino condensates. A more complete understanding of this approach to 
supersymmetry breaking requires coupling higher-derivative supergravity 
to matter and examining the effect of the non-trivial supergravity 
vacuum on the low energy matter Lagrangian. It is clear that 
supersymmetry will be broken in this Lagrangian, but the details of the 
pattern and strength of this breaking require careful study. This study 
is presently underway \cite{PLB-381-154}.

\section*{Acknowledgments}

This work was supported in part by DOE Grant No.\ DE-FG02-95ER40893 and 
NATO Grand No.\ CRG-940784.


\begin{thebibliography}{10}

\bibitem{PRD-53-5583}
A.~Hindawi, B.~A. Ovrut, and D.~Waldram, \emph{Consistent spin-two coupling and
quadratic gravitation}, Phys. Rev. D \textbf{53} (1996), 5583--5596,
hep-th/9509142.

\bibitem{PRD-53-5597}
A.~Hindawi, B.~A. Ovrut, and D.~Waldram, \emph{Non-trivial vacua in higher-
derivative gravitation}, Phys. Rev. D \textbf{53} (1996), 5597--5608, 
hep-th/9509147.

\bibitem{NPB-471-409}
A.~Hindawi, B.~A. Ovrut, and D.~Waldram, \emph{Two-dimensional higher-derivative 
supergravity and a new mechanism for supersymmetry breaking}, Nucl. Phys. 
\textbf{B471} (1996), 409--429, hep-th/9509174.

\bibitem{PLB-133-61}
E.~Cremmer, S.~Ferrara, C.~Kounnas, and D.~V. Nanopoulos, \emph{{Naturally
vanishing cosmological constant in $N=1$ supergravity}}, Phys. Lett.
\textbf{133B} (1983), 61--66.

\bibitem{PP-KFKI-93}
J.~Poloyni, Budapest Preprint KFKI-93 (1977).

\bibitem{PLB-388-512}
K.~Foerger, B.~A. Ovrut, S.~Theisen, and D.~Waldram, \emph{Higher derivative
gravity in string theory}, Phys. Lett. \textbf{388B} (1996) 512--520,
hep-th/9605145.

\bibitem{WB-SS}
J.~Wess and J.~Bagger, \emph{Supersymmetry and supergravity}, 2nd ed.,
Princeton University Press, Princeton, 1992.

\bibitem{NPB-147-105}
E.~Cremmer, B.~Julia, J.~Scherk, S.~Ferrara, L.~Girardello, and P.~van
Nieuwenhuizen, \emph{Spontaneous symmetry breaking and {H}iggs effect in
supergravity without cosmological constant}, Nucl. Phys. \textbf{B147}
(1979), 105--131.

\bibitem{GRG-19-465}
G.~Magnano, M.~Ferraris, and M.~Francaviglia, \emph{Nonlinear gravitational
{L}agrangians}, Gen. Rel. Grav. \textbf{19} (1987), 465--479.

\bibitem{PRD-37-1406}
A.~Jakubiec and J.~Kijowski, \emph{On theories of gravitation with nonlinear
{L}agrangians}, Phys. Rev. D \textbf{37} (1988), 1406--1409.

\bibitem{CQG-5-L95}
M.~Ferraris, M.~Francaviglia, and G.~Magnano, \emph{Do nonlinear metric
  theories of gravitation really exist?}, Class. Quantum Grav. \textbf{5}
  (1988), L95--L99.

\bibitem{CQG-7-557}
G.~Magnano, M.~Ferraris, and M.~Francaviglia, \emph{Legendre transformation and
  dynamical structure of higher-derivative gravity}, Class. Quantum Grav.
  \textbf{7} (1990), 557--570.

\bibitem{NPB-138-430}
S.~Ferrara, M.~T. Grisaru, and P.~van Nieuwenhuizen, \emph{{P}oincar\'e and
  conformal supergravity models with closed algebras}, Nucl. Phys.
  \textbf{B138} (1978), 430--444.

\bibitem{PLB-190-86}
S.~Cecotti, \emph{Higher derivative supergravity is equivalent to standard
  supergravity coupled to matter}, Phys. Lett. \textbf{190B} (1987), 86--92.

\bibitem{PLB-254-132}
B.~A. Ovrut and S.~Kalyana Rama, \emph{{L}orentz and ${U}(1)$ {C}hern-{S}imons
  terms in new minimal supergravity}, Phys. Lett. \textbf{254B} (1991),
  132--138.

\bibitem{NPB-96-331}
L.~O'Raifeartaigh, \emph{Spontaneous symmetry breaking for chiral scalar
  superfields}, Nucl. Phys. \textbf{B96} (1975), 331.

\bibitem{PLB-381-154}
A.~Hindawi, B.~A. Ovrut, and D.~Waldram, \emph{Soft supersymmetry breaking
  induced by higher-derivative supergravitation in the electroweak standard
  model}, Phys. Lett. \textbf{381B} (1996), 154--162, hep-th/9602075.

\end{thebibliography}
\end{document}